\documentclass[12pt]{article}
\usepackage{amssymb}
\usepackage{amsmath}
\allowdisplaybreaks

\begin{document}

\author{C. Bizdadea\thanks{%
e-mail address: bizdadea@central.ucv.ro}, C. C. Ciob\^{\i}rc\u{a}\thanks{%
e-mail address: ciobarca@central.ucv.ro}, E. M. Cioroianu\thanks{%
e-mail address: manache@central.ucv.ro}, S. O. Saliu\thanks{%
e-mail address: osaliu@central.ucv.ro}, \\
Faculty of Physics, University of Craiova\\
13 A. I. Cuza Str., Craiova 200585, Romania}
\title{Interactions between a massless tensor field with the mixed symmetry
of the Riemann tensor and a massless vector field}
\maketitle

\begin{abstract}
Consistent couplings between a massless tensor field with the mixed
symmetry of the Riemann tensor and a massless vector field are
analyzed in the framework of Lagrangian BRST cohomology. Under the
assumptions on smoothness, locality, Lorentz covariance, and
Poincar\'{e} invariance of the deformations, combined with the
requirement that the interacting Lagrangian is at most second-order
derivative, it is proved that there are no consistent
cross-interactions between a single massless tensor field with the
mixed symmetry of the Riemann tensor and one massless vector field.

\end{abstract}

\section{Introduction}

Mixed symmetry type tensor fields~\cite{curt1}--\cite{curt7} are
involved in many physically interesting theories, like superstrings,
supergravities, or supersymmetric high spin theories. The study of
gauge theories with mixed symmetry type tensor fields revealed
several issues, like the dual formulation of field theories of spin
two or higher~\cite{dualsp1}--\cite{dualsp8}, the impossibility of
consistent interactions in the dual formulation of linearized
gravity~\cite{lingr}, or a Lagrangian first-order
approach~\cite{dualsp5,zinov2,zinov3} to some classes of free
massless mixed symmetry type tensor gauge fields, suggestively
resembling to the tetrad formalism of General Relativity. One of the
most important aspects related to this type of gauge models is the
analysis of their consistent interactions, among themselves as well
as with higher-spin gauge theories~\cite{high1}--\cite{4}. The best
approach to this matter is the cohomological one, based on the
deformation of the solution to the master equation~\cite{def}. The
aim of our paper is to investigate the manifestly covariant
consistent interactions between a single, free, massless tensor
gauge field $t_{\mu \nu \vert \alpha \beta }$ with the mixed
symmetry of the Riemann tensor and a massless vector field.

Our procedure relies on the deformation of the solution to the
master equation by means of local BRST cohomology. For each
situation, we initially determine the associated free antifield-BRST
symmetry $s$, which splits as the sum between the Koszul-Tate
differential and the exterior longitudinal derivative only,
$s=\delta +\gamma $. Then, we solve the basic equations of the
deformation procedure. Under the supplementary assumptions on
smoothness, locality, Lorentz covariance, and Poincar\'{e}
invariance of the deformations as well as on the maximum derivative
order of the interacting Lagrangian being equal to two, we prove
that there are no consistent cross-interactions between the tensor
field with the mixed symmetry of the Riemann tensor and the massless
vector field.

The paper is organized in four sections. Section 2 is focused on the
presentation of the free model under study and on the construction
of the associated BRST differential. In Section 3 we briefly review
the antifield-BRST deformation procedure. Section 4 analyzes the
consistent couplings between the tensor field with the mixed
symmetry of the Riemann tensor and the massless vector field with
the help of the local BRST cohomology of the free model. Section 5
ends the paper with some conclusions.

\section{Free model. Free BRST symmetry}

The starting point is given by the free Lagrangian action
\begin{eqnarray}
S_{0}\left[ t_{\mu \nu \vert \alpha \beta },A_{\mu }\right]  &=&\int
d^{D}x\left[ \frac{1}{8}\left( \partial ^{\lambda }t^{\mu \nu \vert
\alpha \beta }\right) \left( \partial _{\lambda }t_{\mu \nu \vert
\alpha \beta }\right) -\left(
\partial _{\mu }t^{\mu \nu \vert \alpha \beta }\right) \left( \partial _{\beta
}t_{\nu \alpha }\right) \right.   \notag \\
&&-\frac{1}{2}\left( \partial _{\mu }t^{\mu \nu \vert \alpha \beta
}\right)
\left( \partial ^{\lambda }t_{\lambda \nu \vert \alpha \beta }\right) -\frac{1}{2}%
\left( \partial ^{\lambda }t^{\nu \beta }\right) \left( \partial _{\lambda
}t_{\nu \beta }\right)   \notag \\
&&+\left( \partial _{\nu }t^{\nu \beta }\right) \left( \partial ^{\lambda
}t_{\lambda \beta }\right) -\frac{1}{2}\left( \partial _{\nu }t^{\nu \beta
}\right) \left( \partial _{\beta }t\right)   \notag \\
&&\left. +\frac{1}{8}\left( \partial ^{\lambda }t\right) \left( \partial
_{\lambda }t\right) -\frac{1}{4}F_{\mu \nu }F^{\mu \nu }\right] \equiv
S_{0}^{\mathrm{t}}\left[ t_{\mu \nu \vert \alpha \beta }\right] +S_{0}^{\mathrm{A}%
}\left[ A_{\mu }\right] ,  \label{r1}
\end{eqnarray}%
in a Minkowski-flat space-time of dimension $D\geq 5$, endowed with
a metric tensor of `mostly plus' signature $\sigma _{\mu \nu
}=\sigma ^{\mu \nu }=\left( -++++\cdots \right) $. The massless
tensor field $t_{\mu \nu \vert \alpha \beta }$ of rank four has the
mixed symmetry of the Riemann tensor and hence transforms according
to an irreducible representation of $GL\left( D,\mathbb{R}\right) $
corresponding to a rectangular Young diagram with two columns and
two rows. Thus, it is separately antisymmetric in the pairs $\left\{
\mu ,\nu \right\} $ and $\left\{ \alpha ,\beta \right\} $, is
symmetric under the interchange of these pairs ($\left\{ \mu ,\nu
\right\} \longleftrightarrow \left\{ \alpha ,\beta \right\} $), and
satisfies the identity $t_{\left[ \mu \nu \vert \alpha \right] \beta
}\equiv 0$ associated with the above diagram, which we will refer to
as the Bianchi I identity. Here and in the sequel the symbol $\left[
\mu \nu \cdots \right] $ signifies complete antisymmetry with
respect to the (Lorentz) indices between brackets, with the
conventions that the minimum number of terms is always used and the
result is never divided by the number of terms. (For instance, we
have that $t_{\left[ \mu \nu \vert \alpha \right] \beta }=t_{\mu \nu
\vert \alpha \beta }+t_{\nu \alpha \vert \mu \beta }+t_{\alpha \mu
\vert \nu \beta }$.) The notation $t_{\nu \beta }$ signifies the
simple trace of the original tensor field, $t_{\nu \beta }=\sigma
^{\mu \alpha }t_{\mu \nu \vert \alpha \beta }$, which is symmetric,
$t_{\nu \beta }=t_{\beta \nu }$, while $t$ denotes its double trace,
$t=\sigma ^{\nu \beta }t_{\nu \beta }\equiv t_{\;\;\;\;\;\vert \mu
\nu }^{\mu \nu }$,\ which is a scalar. A generating set of gauge
transformations for the action (\ref{r1}) reads as
\begin{equation}
\delta _{\epsilon }t_{\mu \nu \vert \alpha \beta }=\partial _{\mu
}\epsilon _{\alpha \beta \vert \nu }-\partial _{\nu }\epsilon
_{\alpha \beta \vert \mu }+\partial _{\alpha }\epsilon _{\mu \nu
\vert \beta }-\partial _{\beta }\epsilon _{\mu \nu \vert \alpha
},\quad \delta _{\epsilon }A_{\mu }=\partial _{\mu }\epsilon ,
\label{r8}
\end{equation}%
with the bosonic gauge parameters $\epsilon _{\mu \nu \vert \alpha
}$
transforming according to an irreducible representation of $GL\left( D,%
\mathbb{R}\right) $ corresponding to a three-cell Young diagram with
two columns and two rows (also known as a hook diagram), being
therefore antisymmetric in the pair $\mu \nu $ and satisfying the
identity $\epsilon _{\left[ \mu \nu \vert \alpha \right] }\equiv 0$.
The last identity is required in order to ensure that the gauge
transformations (\ref{r8}) check the same Bianchi I identity like
the fields themselves, namely, $\delta _{\epsilon }t_{\left[ \mu \nu
\vert \alpha \right] \beta }\equiv 0$. The above generating set of
gauge transformations is Abelian and off-shell, first-stage
reducible since if we make the transformation
\begin{equation}
\epsilon _{\mu \nu \vert \alpha }=2\partial _{\alpha }\theta _{\mu
\nu }-\partial _{\left[ \mu \right. }\theta _{\left. \nu \right]
\alpha },  \label{r12}
\end{equation}%
with $\theta _{\mu \nu }$ an arbitrary antisymmetric tensor ($\theta
_{\mu \nu }=-\theta _{\nu \mu }$), then the gauge transformations of
the tensor field identically vanish, $\delta _{\epsilon \left(
\theta \right) }t_{\mu \nu \vert \alpha \beta }\equiv 0$.

In agreement with the general setting of the antibracket-antifield
formalism, the construction of the BRST symmetry for the free theory under
consideration starts with the identification of the BRST algebra on which
the BRST differential $s$ acts. The generators of the BRST algebra are of
two kinds: fields/ghosts and antifields. The ghost spectrum for the model
under study comprises the fermionic ghosts $\eta _{\alpha \beta \vert \mu }$ and $%
\eta $ associated with the gauge parameters $\epsilon _{\alpha \beta
\vert \mu }$
and $\epsilon $ from (\ref{r8}) as well as the bosonic ghosts for ghosts $%
C_{\mu \nu }$ due to the first-stage reducibility parameters $\theta _{\mu
\nu }$ in (\ref{r12}). In order to make compatible the behaviour of $%
\epsilon _{\alpha \beta \vert \mu }$ and $\theta _{\mu \nu }$ with
that of the corresponding ghosts, we ask that $\eta _{\alpha \beta
\vert \mu }$ satisfies the
properties $\eta _{\mu \nu \vert \alpha }=-\eta _{\nu \mu \vert \alpha }$, $\eta _{%
\left[ \mu \nu \vert \alpha \right] }\equiv 0$ and that $C_{\mu \nu
}$ is
antisymmetric. The antifield spectrum is organized into the antifields $%
t^{\ast \mu \nu \vert \alpha \beta }$ and $A^{\ast \mu }$ of the
original tensor fields and those of the ghosts, $\eta ^{\ast \mu \nu
\vert \alpha }$, $\eta ^{\ast }$, and $C^{\ast \mu \nu }$, of
statistics opposite to that of the associated fields/ghosts. It is
understood that $t^{\ast \mu \nu \vert \alpha \beta }$ is subject to
the conditions
\begin{equation}
t^{\ast \mu \nu \vert \alpha \beta }=-t^{\ast \nu \mu \vert \alpha
\beta }=-t^{\ast \mu \nu \vert \beta \alpha }=t^{\ast \alpha \beta
\vert \mu \nu },\quad t^{\ast \left[ \mu \nu \vert \alpha \right]
\beta }\equiv 0 \label{r43}
\end{equation}%
and, along the same line, it is required that
\begin{equation}
\eta ^{\ast \mu \nu \vert \alpha }=-\eta ^{\ast \nu \mu \vert \alpha
},\quad \eta ^{\ast \left[ \mu \nu \vert \alpha \right] }\equiv
0,\quad C^{\ast \mu \nu }=-C^{\ast \nu \mu }.  \label{r44}
\end{equation}%
We will denote the simple and double traces of $t^{\ast \mu \nu
\vert \alpha \beta }$ by
\begin{equation}
t^{\ast \nu \beta }=\sigma _{\mu \alpha }t^{\ast \mu \nu \vert
\alpha \beta },\quad t^{\ast }=\sigma _{\nu \beta }t^{\ast \nu \beta
},  \label{r44a}
\end{equation}%
such that $t^{\ast \nu \beta }$ is symmetric and $t^{\ast }$ is a scalar.

As both the gauge generators and reducibility functions for this model are
field-independent, it follows that the associated BRST differential ($%
s^{2}=0 $) splits into
\begin{equation}
s=\delta +\gamma ,  \label{r45}
\end{equation}%
where $\delta $ represents the Koszul-Tate differential ($\delta
^{2}=0$), graded by the antighost number $\mathrm{agh}$
($\mathrm{agh}\left( \delta \right) =-1$), and $\gamma $ stands for
the exterior derivative along the gauge orbits. It turns out to be a
true differential ($\gamma ^{2}=0$) that anticommutes with $\delta $
($\delta \gamma +\gamma \delta =0$), whose degree is named pure
ghost number $\mathrm{pgh}$ ($\mathrm{pgh}\left( \gamma \right)
=1$). These two degrees do not interfere ($\mathrm{agh}\left( \gamma
\right) =0$, $\mathrm{pgh}\left( \delta \right) =0$). The overall
degree
that grades the BRST differential is known as the ghost number ($\mathrm{gh}$%
) and is defined like the difference between the pure ghost number and the
antighost number, such that $\mathrm{gh}\left( s\right) =\mathrm{gh}\left(
\delta \right) =\mathrm{gh}\left( \gamma \right) =1$. According to the
standard rules of the BRST method, the corresponding degrees of the
generators from the BRST complex are valued like
\begin{eqnarray}
\mathrm{pgh}\left( t_{\mu \nu \vert \alpha \beta }\right)
&=&0=\mathrm{pgh}\left(
A_{\mu }\right) ,  \label{r46} \\
\mathrm{pgh}\left( \eta _{\mu \nu \vert \alpha }\right)
&=&1=\mathrm{pgh}\left( \eta \right) ,\quad \mathrm{pgh}\left(
C_{\mu \nu }\right) =2,  \label{r47}
\\
\mathrm{pgh}\left( t^{\ast \mu \nu \vert \alpha \beta }\right) &=&\mathrm{pgh}%
\left( A^{\ast \mu }\right) =\mathrm{pgh}\left( \eta ^{\ast \mu \nu
\vert \alpha }\right) =\mathrm{pgh}\left( \eta ^{\ast }\right)
=\mathrm{pgh}\left(
C^{\ast \mu \nu }\right) =0,  \label{r47a} \\
\mathrm{agh}\left( t_{\mu \nu \vert \alpha \beta }\right)
&=&\mathrm{agh}\left(
A_{\mu }\right) =\mathrm{agh}\left( \eta _{\mu \nu \vert \alpha }\right) =\mathrm{%
agh}\left( \eta \right) =\mathrm{agh}\left( C_{\mu \nu }\right) =0,
\label{r48} \\
\mathrm{agh}\left( t^{\ast \mu \nu \vert \alpha \beta }\right) &=&1=\mathrm{agh}%
\left( A^{\ast \mu }\right) ,  \label{r49} \\
\mathrm{agh}\left( \eta ^{\ast \mu \nu \vert \alpha }\right) &=&2=\mathrm{agh}%
\left( \eta ^{\ast }\right) ,\quad \mathrm{agh}\left( C^{\ast \mu \nu
}\right) =3,  \label{r49a}
\end{eqnarray}%
and the actions of $\delta $ and $\gamma $ on them are given by
\begin{eqnarray}
\gamma t_{\mu \nu \vert \alpha \beta } &=&\partial _{\mu }\eta
_{\alpha \beta \vert \nu }-\partial _{\nu }\eta _{\alpha \beta \vert
\mu }+\partial _{\alpha }\eta _{\mu \nu \vert \beta }-\partial
_{\beta }\eta _{\mu \nu \vert \alpha },  \label{r50}
\\
\gamma \eta _{\mu \nu \vert \alpha } &=&2\partial _{\alpha }C_{\mu
\nu }-\partial _{\left[ \mu \right. }C_{\left. \nu \right] \alpha
},\quad \gamma C_{\mu \nu
}=0,  \label{r51} \\
\gamma t^{\ast \mu \nu \vert \alpha \beta } &=&\gamma \eta ^{\ast
\mu \nu \vert \alpha
}=\gamma C^{\ast \mu \nu }=0,  \label{r52} \\
\delta t_{\mu \nu \vert \alpha \beta } &=&\delta \eta _{\mu \nu
\vert \alpha }=\delta
C_{\mu \nu }=0,  \label{r53} \\
\delta t^{\ast \mu \nu \vert \alpha \beta } &=&\frac{1}{4}T^{\mu \nu
\vert \alpha \beta },\quad \delta \eta ^{\ast \alpha \beta \vert \nu
}=-4\partial _{\mu }t^{\ast \mu \nu \vert \alpha \beta },\quad
\delta C^{\ast \mu \nu }=3\partial
_{\alpha }\eta ^{\ast \mu \nu \vert \alpha },  \label{r54} \\
\gamma A_{\mu } &=&\partial _{\mu }\eta ,\quad \gamma \eta =0,\quad \gamma
A^{\ast \mu }=0,\quad \gamma \eta ^{\ast }=0,  \label{r55b} \\
\delta A_{\mu } &=&0,\quad \delta \eta =0,\quad \delta A^{\ast \mu
}=-\partial _{\nu }F^{\mu \nu },\quad \delta \eta ^{\ast }=-\partial _{\mu
}A^{\ast \mu },  \label{r55a}
\end{eqnarray}%
with $T_{\mu \nu \vert \alpha \beta }$ of the form
\begin{eqnarray}
T_{\mu \nu \vert \alpha \beta } &=&\Box t_{\mu \nu \vert \alpha
\beta }+\partial ^{\rho }\left( \partial _{\mu }t_{\alpha \beta
\vert \nu \rho }-\partial _{\nu }t_{\alpha \beta \vert \mu \rho
}+\partial _{\alpha }t_{\mu \nu \vert \beta \rho
}-\partial _{\beta }t_{\mu \nu \vert \alpha \rho }\right)  \notag \\
&&+\left( \partial _{\mu }\partial _{\alpha }t_{\beta \nu }-\partial _{\mu
}\partial _{\beta }t_{\alpha \nu }-\partial _{\nu }\partial _{\alpha
}t_{\beta \mu }+\partial _{\nu }\partial _{\beta }t_{\alpha \mu }\right)
\notag \\
&&-\frac{1}{2}\partial ^{\lambda }\partial ^{\rho }\left( \sigma
_{\mu \alpha }\left( t_{\lambda \beta \vert \nu \rho }+t_{\lambda
\nu \vert \beta \rho }\right) -\sigma _{\mu \beta }\left( t_{\lambda
\alpha \vert \nu \rho
}+t_{\lambda \nu \vert \alpha \rho }\right) \right.  \notag \\
&&\left. -\sigma _{\nu \alpha }\left( t_{\lambda \beta \vert \mu
\rho }+t_{\lambda \mu \vert \beta \rho }\right) +\sigma _{\nu \beta
}\left( t_{\lambda
\alpha \vert \mu \rho }+t_{\lambda \mu \vert \alpha \rho }\right) \right)  \notag \\
&&-\Box \left( \sigma _{\mu \alpha }t_{\beta \nu }-\sigma _{\mu \beta
}t_{\alpha \nu }-\sigma _{\nu \alpha }t_{\beta \mu }+\sigma _{\nu \beta
}t_{\alpha \mu }\right)  \notag \\
&&+\partial ^{\rho }\left( \sigma _{\mu \alpha }\left( \partial _{\beta
}t_{\nu \rho }+\partial _{\nu }t_{\beta \rho }\right) -\sigma _{\mu \beta
}\left( \partial _{\alpha }t_{\nu \rho }+\partial _{\nu }t_{\alpha \rho
}\right) \right.  \notag \\
&&\left. -\sigma _{\nu \alpha }\left( \partial _{\beta }t_{\mu \rho
}+\partial _{\mu }t_{\beta \rho }\right) +\sigma _{\nu \beta }\left(
\partial _{\alpha }t_{\mu \rho }+\partial _{\mu }t_{\alpha \rho }\right)
\right)  \notag \\
&&-\frac{1}{2}\left( \sigma _{\mu \alpha }\partial _{\beta }\partial _{\nu
}-\sigma _{\mu \beta }\partial _{\alpha }\partial _{\nu }-\sigma _{\nu
\alpha }\partial _{\beta }\partial _{\mu }+\sigma _{\nu \beta }\partial
_{\alpha }\partial _{\mu }\right) t  \notag \\
&&-\left( \sigma _{\mu \alpha }\sigma _{\nu \beta }-\sigma _{\mu \beta
}\sigma _{\nu \alpha }\right) \left( \partial ^{\lambda }\partial ^{\rho
}t_{\lambda \rho }-\frac{1}{2}\Box t\right) .  \label{r16}
\end{eqnarray}%
Both $\delta $ and $\gamma $ (and implicitly $s$) were taken to act like
right derivations.

The antifield-BRST differential is known to admit a canonical action in a
structure named antibracket and defined by decreeing the fields/ghosts
conjugated with the corresponding antifields, $s\cdot =\left( \cdot
,S\right) $, where $\left( ,\right) $ signifies the antibracket and $S$
denotes the canonical generator of the BRST symmetry. It is a bosonic
functional of ghost number zero involving both the field/ghost and antifield
spectra, which obeys the classical master equation
\begin{equation}
\left( S,S\right) =0.  \label{r56}
\end{equation}%
The classical master equation is equivalent with the second-order nilpotency
of $s$, $s^{2}=0$, while its solution encodes the entire gauge structure of
the associated theory. Taking into account the formulas (\ref{r50})--(\ref%
{r55a}) as well as the actions of $\delta $ and $\gamma $ in canonical form,
we find that the complete solution to the master equation for the model
under study reads as
\begin{eqnarray}
S &=&S_{0}\left[ t_{\mu \nu \vert \alpha \beta },A_{\mu }\right] +\int d^{D}x%
\left[ t^{\ast \mu \nu \vert \alpha \beta }\left( \partial _{\mu
}\eta _{\alpha \beta \vert \nu }-\partial _{\nu }\eta _{\alpha \beta
\vert \mu }+\partial _{\alpha
}\eta _{\mu \nu \vert \beta }\right. \right.  \notag \\
&&\left. \left. -\partial _{\beta }\eta _{\mu \nu \vert \alpha
}\right) +\eta ^{\ast \mu \nu \vert \alpha }\left( 2\partial
_{\alpha }C_{\mu \nu }-\partial _{ \left[ \mu \right. }C_{\left. \nu
\right] \alpha }\right) +A^{\ast \mu }\partial _{\mu }\eta \right] .
\label{r55}
\end{eqnarray}%
The main ingredients of the antifield-BRST symmetry derived in this section
will be useful in the sequel at the analysis of consistent interactions that
can be added to the action (\ref{r1}) without changing its number of
independent gauge symmetries.

\section{Brief review of the antifield-BRST deformation procedure}

There are three main types of consistent interactions that can be
added to a given gauge theory: the first type deforms only the
Lagrangian action, but not its gauge transformations, the second
kind modifies both the action and its transformations, but not the
gauge algebra, and the third, and certainly most interesting
category, changes everything, namely, the action, its gauge
symmetries, and the accompanying algebra.

The reformulation of the problem of consistent deformations of a given
action and of its gauge symmetries in the antifield-BRST setting is based on
the observation that if a deformation of the classical theory can be
consistently constructed, then the solution to the master equation for the
initial theory can be deformed into
\begin{equation}
\bar{S}=S+gS_{1}+g^{2}S_{2}+O\left( g^{3}\right) ,\quad \varepsilon \left(
\bar{S}\right) =0,\quad \mathrm{gh}\left( \bar{S}\right) =0,  \label{r57}
\end{equation}%
such that
\begin{equation}
\left( \bar{S},\bar{S}\right) =0.  \label{r58}
\end{equation}%
Here and in the sequel $\varepsilon \left( F\right) $ denotes the Grassmann
parity of $F$. The projection of (\ref{r58}) on the various powers in the
coupling constant induces the following tower of equations:
\begin{eqnarray}
g^{0} &:&\left( S,S\right) =0,  \label{r59} \\
g^{1} &:&\left( S_{1},S\right) =0,  \label{r60} \\
g^{2} &:&\frac{1}{2}\left( S_{1},S_{1}\right) +\left( S_{2},S\right) =0,
\label{r61} \\
&&\vdots  \notag
\end{eqnarray}%
The first equation is satisfied by hypothesis. The second one
governs the first-order deformation of the solution to the master
equation ($S_{1}$) and it shows that $S_{1}$ is a BRST co-cycle,
$sS_{1}=0$. This means that $S_{1}$ pertains to the ghost number
zero cohomological space of $s$, $H^{0}\left( s\right) $, which is
generically non-empty due to its isomorphism to the space of
physical observables of the free theory. The remaining equations are
responsible for the higher-order deformations of the solution to the
master equation. No obstructions arise in finding solutions to them
as long as no further restrictions, such as space-time locality, are
imposed. Obviously, only nontrivial first-order deformations should
be considered, since trivial ones ($S_{1}=sB$) lead to trivial
deformations of the initial theory and can be eliminated by
convenient redefinitions of the fields. Ignoring the trivial
deformations, it follows that $S_{1}$ is a nontrivial
BRST-observable, $S_{1}\in
H^{0}\left( s\right) $. Once that the deformation equations (\ref{r60})--(%
\ref{r61}), etc., have been solved by means of specific cohomological
techniques, from the consistent nontrivial deformed solution to the master
equation we can extract all the information on the gauge structure of the
accompanying interacting theory.

\section{First-order deformation}

The purpose of our paper is to study the consistent interactions that can be
added to the free action (\ref{r1}) by means of solving the main deformation
equations, namely, (\ref{r60})--(\ref{r61}), etc. For obvious reasons, we
consider only smooth, local, Lorentz-covariant, and Poincar\'{e}-invariant
deformations. If we make the notation $S_{1}=\int d^{D}x\,a$, with $a$ a
local function, then the local form of the equation (\ref{r60}), which we
have seen that controls the first-order deformation of the solution to the
master equation, becomes
\begin{equation}
sa=\partial _{\mu }m^{\mu },\quad \mathrm{gh}\left( a\right) =0,\quad
\varepsilon \left( a\right) =0,  \label{r62}
\end{equation}%
for some $m^{\mu }$, and it shows that the nonintegrated density of
the first-order deformation pertains to the local cohomology of $s$
at ghost number zero, $a\in H^{0}\left( s\vert d\right) $, where $d$
denotes the exterior
space-time differential. In order to analyze the above equation, we develop $%
a$ according to the antighost number
\begin{equation}
a=\sum\limits_{k=0}^{I}a_{k},\quad \mathrm{agh}\left( a_{k}\right) =k,\quad
\mathrm{gh}\left( a_{k}\right) =0,\quad \varepsilon \left( a_{k}\right) =0,
\label{r63}
\end{equation}%
and assume, without loss of generality, that $a$ stops at some finite value $%
I$ of the antighost number.\footnote{%
This can be shown, for instance, like in~\cite{gen2a} (Section 3) or \cite%
{gen2b}, under the sole assumption that the interacting Lagrangian at the
first order in the coupling constant, $a_{0}$, has a finite, but otherwise
arbitrary derivative order.} By taking into account the decomposition (\ref%
{r45}) of the BRST differential, the equation (\ref{r62}) is equivalent to a
tower of local equations, corresponding to the various decreasing values of
the antighost number
\begin{eqnarray}
\gamma a_{I} &=&\partial _{\mu }\overset{(I)}{m}^{\mu },  \label{r64} \\
\delta a_{I}+\gamma a_{I-1} &=&\partial _{\mu }\overset{(I-1)}{m}^{\mu },
\label{r65} \\
\delta a_{k}+\gamma a_{k-1} &=&\partial _{\mu }\overset{(k-1)}{m}^{\mu
},\quad I-1\geq k\geq 1,  \label{r66}
\end{eqnarray}%
where $\left( \overset{(k)}{m}^{\mu }\right) _{k=\overline{0,I}}$ are some
local currents, with $\mathrm{agh}\left( \overset{(k)}{m}^{\mu }\right) =k$.
It can be proved\footnote{%
The fact that it is possible to replace the equation (\ref{r64}) with (\ref%
{r67}) can be done like in the proof of Corollary 3 from~\cite{IJGMMP}, with
the precaution to include in an appropriate manner the dependence on the
vector field BRST sector.} that one can replace the equation (\ref{r64}) at
strictly positive antighost numbers with
\begin{equation}
\gamma a_{I}=0,\quad I>0.  \label{r67}
\end{equation}%
In conclusion, under the assumption that $I>0$, the representative of
highest antighost number from the nonintegrated density of the first-order
deformation can always be taken to be $\gamma $-closed, such that the
equation (\ref{r62}) associated with the local form of the first-order
deformation is completely equivalent to the tower of equations (\ref{r67})
and (\ref{r65})--(\ref{r66}).

Before proceeding to the analysis of the solutions to the first-order
deformation equation, we briefly comment on the uniqueness and triviality of
such solutions. Due to the second-order nilpotency of $\gamma $ ($\gamma
^{2}=0$), the solution to the top equation (\ref{r67}) is clearly unique up
to $\gamma $-exact contributions,
\begin{equation}
a_{I}\rightarrow a_{I}+\gamma b_{I},\quad \mathrm{agh}\left( b_{I}\right)
=I,\quad \mathrm{pgh}\left( b_{I}\right) =I-1,\quad \varepsilon \left(
b_{I}\right) =1.  \label{r68}
\end{equation}%
Meanwhile, if it turns out that $a_{I}$ reduces to $\gamma $-exact terms
only, $a_{I}=\gamma b_{I}$, then it can be made to vanish, $a_{I}=0$. In
other words, the nontriviality of the first-order deformation $a$ is
translated at its highest antighost number component into the requirement
that
\begin{equation}
a_{I}\in H^{I}\left( \gamma \right) ,  \label{r69}
\end{equation}%
where $H^{I}\left( \gamma \right) $ denotes the cohomology of the exterior
longitudinal derivative $\gamma $ at pure ghost number equal to $I$. At the
same time, the general condition on the nonintegrated density of the
first-order deformation to be in a nontrivial cohomological class of $%
H^{0}\left( s\vert d\right) $ shows on the one hand that the solution to (\ref%
{r62}) is unique up to $s$-exact pieces plus total divergences
\begin{equation}
a\rightarrow a+sb+\partial _{\mu }n^{\mu },\quad \mathrm{gh}\left( b\right)
=-1,\quad \varepsilon \left( b\right) =1,\quad \mathrm{gh}\left( n^{\mu
}\right) =0,\quad \varepsilon \left( n^{\mu }\right) =0  \label{r73}
\end{equation}%
and on the other hand that if the general solution to (\ref{r62}) is found
to be completely trivial, $a=sb+\partial _{\mu }n^{\mu }$, then it can be
made to vanish, $a=0$.

\subsection{Basic cohomologies}

In the light of the above discussion, we pass to the investigation of the
solutions to the equations (\ref{r67}) and (\ref{r65})--(\ref{r66}). We have
seen that $a_{I}$ belongs to the cohomology of the exterior longitudinal
derivative (see the formula (\ref{r69})), such that we need to compute $%
H\left( \gamma \right) $ in order to construct the component of highest
antighost number from the first-order deformation. This matter is solved
with the help of the definitions (\ref{r50})--(\ref{r52}) and (\ref{r55b}).

In order to determine the cohomology $H(\gamma )$, we split the differential
$\gamma $ into two pieces
\begin{equation}
\gamma =\gamma _{t}+\gamma _{A},  \label{3.7}
\end{equation}%
where $\gamma _{t}$ acts nontrivially only on the fields/ghosts from the $%
t_{\mu \nu \vert \alpha \beta }$ sector and $\gamma _{A}$ does the
same thing, but with respect to the vector field sector. From the
above splitting it follows that the nilpotency of $\gamma $ is
equivalent to the nilpotency and anticommutativity of its components
\begin{equation}
\left( \gamma _{t}\right) ^{2}=0=\left( \gamma _{A}\right)
^{2},\quad \gamma _{t}\gamma _{A}+\gamma _{A}\gamma _{t}=0.
\label{3.8}
\end{equation}%
Kunneth's formula then ensures the isomorphism
\begin{equation}
H(\gamma )=H(\gamma _{t})\otimes H(\gamma _{A}).  \label{3.9}
\end{equation}%
Thus, we can state that $H(\gamma )$ is generated~\cite{EPJC} on the
one hand by $\chi ^{\ast \Delta }$, $F_{\mu \nu }$, and $F_{\mu \nu
\lambda \vert \alpha \beta \gamma }$ as well as by their space-time
derivatives and on the other hand
by the ghosts $C_{\mu \nu }$, $\partial _{\left[ \mu \right. }C_{\left. \nu %
\right] \alpha }$, and $\eta $, where $\chi ^{\ast \Delta }$ is a collective
notation for all the antifields
\begin{equation}
\chi ^{\ast \Delta }=\left\{ t^{\ast \mu \nu \vert \alpha \beta
},A^{\ast \mu },\eta ^{\ast \mu \nu \vert \alpha },\eta ^{\ast
},C^{\ast \mu \nu }\right\} , \label{r73ab}
\end{equation}%
while
\begin{eqnarray}
F_{\mu \nu \lambda \vert \alpha \beta \gamma } &=&\partial _{\lambda
}\partial _{\gamma }t_{\mu \nu \vert \alpha \beta }+\partial _{\mu
}\partial _{\gamma }t_{\nu \lambda \vert \alpha \beta }+\partial
_{\nu }\partial _{\gamma
}t_{\lambda \mu \vert \alpha \beta }  \notag \\
&&+\partial _{\lambda }\partial _{\alpha }t_{\mu \nu \vert \beta
\gamma }+\partial _{\mu }\partial _{\alpha }t_{\nu \lambda \vert
\beta \gamma }+\partial
_{\nu }\partial _{\alpha }t_{\lambda \mu \vert \beta \gamma }  \notag \\
&&+\partial _{\lambda }\partial _{\beta }t_{\mu \nu \vert \gamma
\alpha }+\partial _{\mu }\partial _{\beta }t_{\nu \lambda \vert
\gamma \alpha }+\partial _{\nu }\partial _{\beta }t_{\lambda \mu
\vert \gamma \alpha },  \label{curvature}
\end{eqnarray}%
represent the components of the curvature tensor for $t_{\mu \nu
\vert \alpha \beta }$ (the quantities with the minimum number of
derivatives, invariant under the gauge transformations $\delta
_{\epsilon }t_{\mu \nu \vert \alpha \beta }$ in (\ref{r8})). (The
quantity $\partial _{\left[ \mu \right. }C_{\left. \nu \right]
\alpha }$ carries a trivial component. Its non-trivial part is given
by the completely antisymmetric expression $\partial _{\left[ \mu
\right. }C_{\left. \nu \alpha \right]}$, which differs from our
representative by a $\gamma $-exact term.) So, the most general (and
nontrivial), local solution to (\ref{r67}) can be written, up to
$\gamma $-exact contributions, as
\begin{equation}
a_{I}=\alpha _{I}\left( \left[ F_{\mu \nu }\right] ,\left[ F_{\mu
\nu \lambda \vert \alpha \beta \gamma }\right] ,\left[ \chi ^{\ast
\Delta }\right] \right) \omega ^{I}\left( C_{\mu \nu },\partial
_{\left[ \mu \right. }C_{\left. \nu \right] \alpha },\eta \right) ,
\label{r79}
\end{equation}%
where the notation $f([q])$ means that $f$ depends on $q$ and its
derivatives up to a finite order and $\omega ^{I}$ denotes the elements of a
basis in the space of polynomials with pure ghost number $I$ in the
corresponding ghosts and some of their first-order derivatives. The objects $%
\alpha _{I}$ (obviously nontrivial in $H^{0}\left( \gamma \right) $)
were taken to have a bounded number of derivatives and therefore
they are polynomials in the antifields $\chi ^{\ast \Delta }$, in
$F_{\mu \nu }$, in the curvature tensor $F_{\mu \nu \lambda \vert
\alpha \beta \gamma }$ as well as in their derivatives. Due to the
fact that these elements are $\gamma $-closed, they are called
invariant polynomials. At zero antighost number, the invariant
polynomials are polynomials in the curvature tensor $F_{\mu \nu
\lambda \vert \alpha \beta \gamma }$, the field strength $F_{\mu \nu
}$, and their derivatives.

Replacing the solution (\ref{r79}) in the equation (\ref{r65}), we
remark that a necessary (but not sufficient) condition for the
existence of (nontrivial) solutions $a_{I-1}$ is that the invariant
polynomials $\alpha _{I}$ from (\ref{r79}) are (nontrivial) objects
from the local cohomology of the Koszul-Tate differential $H\left(
\delta \vert d\right) $ at antighost number $I>0$ and pure ghost
number equal to zero\footnote{\label{footnote}We recall that the
local cohomology $H\left( \delta \vert d\right) $ is completely
trivial at both strictly positive antighost \textit{and} pure ghost
numbers (for instance, see~\cite{gen1a}, Theorem 5.4 or
\cite{gen1b}).}, $\alpha _{I}\in H_{I}\left( \delta \vert d\right)
$, i.e.
\begin{equation}
\delta \alpha _{I}=\partial _{\mu }j^{\mu },\quad \varepsilon \left( j^{\mu
}\right) =1,\quad \mathrm{agh}\left( j^{\mu }\right) =I-1,\quad \mathrm{pgh}%
\left( j^{\mu }\right) =0.  \label{r80}
\end{equation}%
(In view of the footnote \ref{footnote} from now on it is understood
that by $H_{I}\left( \delta \vert d\right) $ we mean the local
cohomology of the Koszul-Tate differential at antighost $I$ and at
pure ghost number zero.) Consequently, we need to investigate some
of the main properties of the local cohomology of the Koszul-Tate
differential at strictly positive antighost numbers in order to
completely determine the component $a_{I}$ of highest antighost
number in the first-order deformation. As the free model under study
is a normal gauge theory of Cauchy order equal to three, the general
results from~\cite{gen1a,gen1b} ensure that the local cohomology of
the Koszul-Tate differential is trivial at antighost numbers
strictly greater than its Cauchy order
\begin{equation}
H_{k}\left( \delta \vert d\right) =0,\quad k>3.  \label{r81}
\end{equation}%
Moreover, if the invariant polynomial $\alpha _{k}$, with $\mathrm{agh}%
\left( \alpha _{k}\right) =k\geq 3$, is trivial in $H_{k}\left( \delta
\vert d\right) $, then it can be taken to be trivial also in $H_{k}^{\mathrm{inv}%
}\left( \delta \vert d\right) $%
\begin{equation}
\left( \alpha _{k}=\delta b_{k+1}+\partial _{\mu }\overset{(k)}{c}^{\mu
},\quad \mathrm{agh}\left( \alpha _{k}\right) =k\geq 3\right) \Rightarrow
\alpha _{k}=\delta \beta _{k+1}+\partial _{\mu }\overset{(k)}{\gamma }^{\mu
},  \label{r81d}
\end{equation}%
where $\beta _{k+1}$ and $\overset{(k)}{\gamma }^{\mu }$ are invariant
polynomials\footnote{%
The proof can be realized in the same manner like Theorem 5 from~\cite%
{IJGMMP}, with the precaution to include in an appropriate manner the
dependence on the vector field BRST sector.}. [An element of $H_{k}^{\mathrm{%
inv}}\left( \delta \vert d\right) $ is defined via an equation similar to (\ref%
{r80}), but with the corresponding current an invariant polynomial.]
The results (\ref{r81d}) and (\ref{r81}) ensure that all the local
cohomology of the Koszul-Tate differential in the space of invariant
polynomials is trivial in antighost numbers strictly greater than
three
\begin{equation}
H_{k}^{\mathrm{inv}}\left( \delta \vert d\right) =0,\quad k>3.
\label{r81c}
\end{equation}%
Using the definitions (\ref{r54}) and (\ref{r55a}), we can organize
the nontrivial representatives of $\left( H_{k}\left( \delta \vert
d\right) \right) _{k\geq 2}$ and $\left( H_{k}^{\mathrm{inv}}\left(
\delta \vert d\right) \right) _{k\geq 2}$ like: i) for $k>3$ there
are none; ii) for $k=3$ they are linear combinations of the
undifferentiated antifields $C^{\ast \mu \nu }$ with constant
coefficients; iii) for $k=2$ they are written like linear
combinations of the undifferentiated antifields $\eta ^{\ast \mu \nu
\vert \alpha }$ and $\eta ^{\ast }$ with constant coefficients. We
have excluded from both $H\left( \delta \vert d\right) $ and
$H^{\mathrm{inv}}\left( \delta \vert d\right) $ the nontrivial
elements depending on the space-time coordinates, as they would
result in interactions with broken Poincar\'{e} invariance. In
contrast to the groups $\left( H_{k}\left( \delta \vert d\right)
\right) _{k\geq 2}$ and $\left( H_{k}^{\mathrm{inv}}\left( \delta
\vert d\right) \right) _{k\geq 2}$, which are finite-dimensional,
the cohomology $H_{1}\left( \delta \vert d\right) $, that is related
to global symmetries and ordinary conservation laws, is
infinite-dimensional since the theory is free. Fortunately, it will
not be needed in the sequel.

The previous results on $H\left( \delta \vert d\right) $ and $H^{\mathrm{inv}%
}\left( \delta \vert d\right) $ at strictly positive antighost
numbers are important because they control the obstructions to
removing the antifields
from the first-order deformation. As a consequence of the result (\ref{r81c}%
), we can eliminate all the terms with $k>3$ from the expansion (\ref{r63})
by adding only trivial pieces and thus work with $I\leq 3$.

\subsection{Computation of the first-order deformation}

Now, we have at hand all the necessary ingredients for computing the general
form of the first-order deformation of the solution to the master equation.
In the case $I=3$ the nonintegrated density of the first-order deformation
becomes
\begin{equation}
a=a_{0}+a_{1}+a_{2}+a_{3}.  \label{3.12}
\end{equation}%
We can further decompose $a$ in a natural manner as
\begin{equation}
a=a^{\mathrm{t}}+a^{\mathrm{t-A}}+a^{\mathrm{A}},  \label{3.12a}
\end{equation}%
where $a^{\mathrm{t}}$ contains only fields/ghosts/antifields from the $%
t_{\mu \nu \vert \alpha \beta }$ sector, $a^{\mathrm{t-A}}$
describes the cross-interactions between the tensor field $t_{\mu
\nu \vert \alpha \beta }$ the vector field (so it effectively mixes
both sectors), and $a^{\mathrm{A}}$ involves only the vector field
sector. As it has been shown in \cite{EPJC} under the hypotheses of
smoothness, locality, Lorentz covariance, and Poincar\'{e}
invariance of the deformations, combined with the requirement that
the interacting Lagrangian is at most second-order derivative, to be
maintained here as well, $a^{\mathrm{t}}$
satisfies an equation similar to (\ref{r62}) and has the expression%
\begin{equation}
a^{\mathrm{t}}=c^{\prime }t\equiv t_{\;\;\;\;\;\vert \mu \nu }^{\mu
\nu }, \label{xz1}
\end{equation}%
with $c^{\prime }$ an arbitrary, real constant. On the other hand $a^{%
\mathrm{t-A}}$ and $a^{\mathrm{A}}$ involve different sorts of
fields, so these components verify independently some equations
similar to (\ref{r62})
\begin{eqnarray}
sa^{\mathrm{t-A}} &=&\partial _{\mu }m^{\left( \mathrm{t-A}\right) \mu },
\label{uw1} \\
sa^{\mathrm{A}} &=&\partial _{\mu }m^{\left( \mathrm{A}\right) \mu },
\label{uw2}
\end{eqnarray}%
for some local $m^{\mu }$'s. In the sequel we analyze the general solutions
to these equations.

The term $a^{\mathrm{t-A}}$ allows a decomposition similar to (\ref{3.12})%
\begin{equation}
a^{\mathrm{t-A}}=a_{0}^{\mathrm{t-A}}+a_{1}^{\mathrm{t-A}}+a_{2}^{\mathrm{t-A%
}}+a_{3}^{\mathrm{t-A}},  \label{3.25}
\end{equation}%
where the components of $a^{\mathrm{t-A}}$ are subject to the equations
\begin{eqnarray}
\gamma a_{3}^{\mathrm{t-A}} &=&0,  \label{3.25q} \\
\delta a_{I}^{\mathrm{t-A}}+\gamma a_{I-1}^{\mathrm{t-A}} &=&\partial _{\mu }%
\overset{\left( I-1\right) }{m}^{\left( \mathrm{t-A}\right) \mu },\quad
I=1,2,3.  \label{3.25w}
\end{eqnarray}%
In agreement with (\ref{r79}) and with the discussion made in the
above regarding the nontrivial representatives of
$H_{3}^{\mathrm{inv}}\left( \delta \vert d\right) $ (see the case
ii)) the equation (\ref{3.25q}) possesses
in $D\geq 5$ space-time dimensions the solution%
\begin{equation}
a_{3}^{\mathrm{t-A}}=a_{3}^{\left( 1\right) \mathrm{t-A}}+a_{3}^{\left(
2\right) \mathrm{t-A}},  \label{x1}
\end{equation}%
where:

\begin{itemize}
\item for all $D\geq 5$%
\begin{equation}
a_{3}^{\left( 1\right) \mathrm{t-A}}=c_{1}C^{\ast \mu \nu }C_{\mu \nu }\eta ;
\label{r134}
\end{equation}

\item for $D=5$%
\begin{equation}
a_{3}^{\left( 2\right) \mathrm{t-A}}=c_{2}\varepsilon ^{\mu \nu \lambda
\beta \rho }C_{\mu \nu }^{\ast }\left( \partial _{\lambda }C_{\beta \rho
}\right) \eta .  \label{r133}
\end{equation}
\end{itemize}

In the relations (\ref{r134})--(\ref{r133}) $c_{1}$ and $c_{2}$ are some
arbitrary, real constants. Obviously, since the components (\ref{r134})--(%
\ref{r133}) are mutually independent, it follows that each of them must
separately fulfill an equation of type (\ref{3.25w}) for $I=3$
\begin{equation}
\delta a_{3}^{(i)\mathrm{t-A}}=-\gamma a_{2}^{(i)\mathrm{t-A}}+\partial
_{\mu }\overset{\left( 2\right) }{m}^{\left( i\right) \left( \mathrm{t-A}%
\right) \mu },\quad i=1,2.  \label{r135}
\end{equation}%
By direct computation we obtain
\begin{eqnarray}
\delta a_{3}^{\left( 1\right) \mathrm{t-A}} &=&-\gamma \left[
c_{1}\eta ^{\ast \mu \nu \vert \alpha }\left( \frac{3}{2}\eta _{\mu
\nu \vert \alpha }\eta
-C_{\mu \nu }A_{\alpha }\right) \right] +\partial _{\mu }u^{\mu }  \notag \\
&&+\frac{3}{2}c_{1}\eta ^{\ast \mu \nu \vert \alpha }\partial
_{\left[ \mu \right. }C_{\left. \nu \right] \alpha }\eta .
\label{x136}
\end{eqnarray}%
Thus, $a_{3}^{\left( 1\right) \mathrm{t-A}}$ produces a consistent $%
a_{2}^{\left( 1\right) \mathrm{t-A}}$ as solution to the equation (\ref{r135}%
) for $i=1$ if and only if the term $\left( 3/2\right) c_{1}\eta
^{\ast \mu \nu \vert \alpha }\partial _{\left[ \mu \right.
}C_{\left. \nu \right] \alpha }\eta $, which is a nontrivial
representative of $H\left( \gamma \right) $, is written in a $\gamma
$-exact modulo $d$ form. This takes place if and
only if%
\begin{equation}
c_{1}=0.  \label{x138}
\end{equation}%
Related to $a_{3}^{\left( 2\right) \mathrm{t-A}}$, by applying $\delta $ on (%
\ref{r133}) we find via the equation (\ref{r135}) for $i=2$ that $%
a_{2}^{\left( 2\right) \mathrm{t-A}}$ reads as
\begin{eqnarray}
a_{2}^{\left( 2\right) \mathrm{t-A}} &=&3c_{2}\varepsilon ^{\mu \nu \lambda
\beta \rho }\eta _{\mu \nu \vert }^{\ast \;\;\;\alpha }\left[ \left( \frac{5}{12}%
\partial _{\lambda }\eta _{\beta \rho \vert \alpha }+\frac{1}{6}\partial _{\beta
}\eta _{\rho \alpha \vert \lambda }\right) \eta \right.  \notag \\
&&\left. -\left( \partial _{\lambda }C_{\beta \rho }\right) A_{\alpha }
\right] .  \label{x139}
\end{eqnarray}%
By means of (\ref{x139}) we deduce
\begin{eqnarray}
\delta a_{2}^{\left( 2\right) \mathrm{t-A}} &=&\partial _{\mu
}v^{\mu }-\gamma \left[ c_{2}\varepsilon ^{\mu \nu \lambda \beta
\rho }t_{\;\;\;\;\;\;\;\mu \nu }^{\ast \tau \alpha \vert }\left(
3\left( \partial _{\lambda }t_{\tau \alpha \vert \beta \rho }\right)
\eta \right. \right.  \notag
\\
&&\left. \left. +\left( 5\partial _{\lambda }\eta _{\beta \rho \vert
\tau }+2\partial _{\beta }\eta _{\rho \tau \vert \lambda }\right)
A_{\alpha }\right)
\right]  \notag \\
&&-6c_{2}\varepsilon ^{\mu \nu \lambda \beta \rho
}t_{\;\;\;\;\;\;\;\mu \nu }^{\ast \tau \alpha \vert }\left( \partial
_{\lambda }C_{\beta \rho }\right) F_{\tau \alpha }.  \label{x140}
\end{eqnarray}%
Comparing (\ref{x140}) with (\ref{3.25w}) for $I=2$ we can state that $%
a_{2}^{\left( 2\right) \mathrm{t-A}}$ provides a consistent
$a_{1}^{\left( 2\right) \mathrm{t-A}}$ if and only if the term
$6c_{2}\varepsilon ^{\mu \nu \lambda \beta \rho
}t_{\;\;\;\;\;\;\;\mu \nu }^{\ast \tau \alpha \vert }\left(
\partial _{\lambda }C_{\beta \rho }\right) F_{\tau \alpha }$, which is again
a\ nontrivial representative of $H\left( \gamma \right) $, is $\gamma $%
-exact modulo $d$. This holds if and only if%
\begin{equation}
c_{2}=0.  \label{x141}
\end{equation}%
The results (\ref{x138}) and (\ref{x141}) show that%
\begin{equation}
a_{3}^{\mathrm{t-A}}=0.  \label{x142}
\end{equation}

Accordingly, $a^{\mathrm{t-A}}$ can stop earliest at antighost number two%
\begin{equation}
a^{\mathrm{t-A}}=a_{0}^{\mathrm{t-A}}+a_{1}^{\mathrm{t-A}}+a_{2}^{\mathrm{t-A%
}},  \label{xtv91}
\end{equation}%
with $a_{2}^{\mathrm{t-A}}$ solution to the equation $\gamma a_{2}^{\mathrm{%
t-A}}=0$. Looking at (\ref{r79}) for $I=2$, using the previous results on
the nontrivial representatives of $H_{2}^{\mathrm{inv}}\left( \delta
\vert d\right) $ (see the case iii) in the above), and requiring that $a_{2}^{%
\mathrm{t-A}}$ effectively describes cross-couplings, we get (up to trivial,
$\gamma $-exact contributions) that%
\begin{equation}
a_{2}^{\mathrm{t-A}}=\eta ^{\ast }\left( \lambda ^{\mu \nu }C_{\mu \nu }+%
\bar{\lambda}^{\mu \nu \alpha }\partial _{\left[ \mu \right. }C_{\left. \nu %
\right] \alpha }\right) ,  \label{x144}
\end{equation}%
where $\lambda ^{\mu \nu }$ and $\bar{\lambda}^{\mu \nu \alpha }$
are some non-derivative, real constant tensors, invariant under the
Lorentz group, with $\lambda ^{\mu \nu }$ and $\bar{\lambda}%
^{\mu \nu \alpha }$ antisymmetric in $\lambda $ and $\mu $. Since in
$D\geq
5 $ there are no such constants we must set $\lambda ^{\mu \nu }=0$ and $%
\bar{\lambda}^{\mu \nu \alpha }=0$, so we have that%
\begin{equation}
a_{2}^{\mathrm{t-A}}=0.  \label{x143}
\end{equation}

In this way we infer that $a^{\mathrm{t-A}}$ actually stops at antighost
number one%
\begin{equation}
a^{\mathrm{t-A}}=a_{0}^{\mathrm{t-A}}+a_{1}^{\mathrm{t-A}},  \label{xtv98}
\end{equation}%
with $a_{1}^{\mathrm{t-A}}$ solution to the equation $\gamma a_{1}^{\mathrm{%
t-A}}=0$. Since $a_{1}^{\mathrm{t-A}}$ is linear in the antifields of the
original fields, we can write
\begin{equation}
a_{1}^{\mathrm{t-A}}=\left( t_{\mu \nu \vert \alpha \beta }^{\ast
}\Delta ^{\mu \nu \vert \alpha \beta }+A_{\mu }^{\ast }\Delta ^{\mu
}\right) \eta , \label{x145}
\end{equation}%
where $\Delta ^{\mu \nu \vert \alpha \beta }$ and $\Delta ^{\mu }$ are $\gamma $%
-invariant objects (with both the antighost number and the pure
ghost number equal to zero), the former quantities displaying the
mixed symmetry of the tensor $t^{\mu \nu \vert \alpha \beta }$. From
(\ref{r79}) at antighost number zero we observe that $\Delta ^{\mu
\nu \vert \alpha \beta }$ and $\Delta ^{\mu }$ depend in general on
$F^{\mu \nu }$, $F^{\mu \nu \lambda \vert \alpha \beta \gamma }$,
and their derivatives. Moreover, the requirement that the second
term in the right-hand side of (\ref{x145}) produces
cross-interactions implies that $\Delta ^{\mu }$ must involve
$F^{\mu \nu \lambda \vert \alpha \beta
\gamma }$ (and possibly its derivatives). In order to construct $a_{0}^{%
\mathrm{t-A}}$ as solution to the equation%
\begin{equation}
\delta a_{1}^{\mathrm{t-A}}+\gamma a_{0}^{\mathrm{t-A}}=\partial _{\mu }%
\overset{\left( 0\right) }{m}^{\left( \mathrm{t-A}\right) \mu },
\label{x146}
\end{equation}%
we invoke the hypothesis on the maximum derivative order of $a_{0}^{\mathrm{%
t-A}}$ being equal to two. As both $\delta t_{\mu \nu \vert \alpha
\beta }^{\ast } $ and $\delta A_{\mu }^{\ast }$ contain exactly two
derivatives, it follows that each of $\Delta ^{\mu \nu \vert \alpha
\beta }$ and $\Delta ^{\mu }$ are allowed to include at most one
derivative (we remind that $\partial
_{\mu }\eta =\gamma A_{\mu }$) and therefore we have that%
\begin{equation}
\Delta ^{\mu \nu \vert \alpha \beta }=\Delta _{0}^{\mu \nu \vert
\alpha \beta }+\Delta _{1}^{\mu \nu \vert \alpha \beta },\quad
\Delta ^{\mu }=\Delta _{0}^{\mu }+\Delta _{1}^{\mu },  \label{x147}
\end{equation}%
where $\Delta _{0}$'s contain no derivatives and $\Delta _{1}$'s include
just one derivative. Therefore, both $\Delta _{0}$'s must be constant, while
$\Delta _{1}$'s must depend linearly on $F_{\mu \nu }$. From covariance
arguments in $D\geq 5$ we have that the only possible choice of these
quantities is%
\begin{equation}
\Delta _{0}^{\mu \nu \vert \alpha \beta }=\frac{1}{2}k_{1}\left(
\sigma ^{\mu \alpha }\sigma ^{\nu \beta }-\sigma ^{\mu \beta }\sigma
^{\nu \alpha }\right) ,\quad \Delta _{0}^{\mu }=0,  \label{x148}
\end{equation}%
and
\begin{equation}
\Delta _{1}^{\mu \nu \vert \alpha \beta }=k_{2}\varepsilon ^{\mu \nu
\alpha \beta \lambda \rho }F_{\lambda \rho },\quad \Delta _{1}^{\mu
}=0,  \label{x149}
\end{equation}%
with $k_{1}$ and $k_{2}$ arbitrary, real constants. The solution (\ref{x149}%
) `lives' in $D=6$, but it brings no contribution to
$a_{1}^{\mathrm{t-A}}$ as $\varepsilon ^{\mu \nu \alpha \beta
\lambda \rho }t_{\mu \nu \vert \alpha
\beta }^{\ast }\equiv 0$. Substituting (\ref{x147})--(\ref{x149}) in (\ref%
{x145}) we find that%
\begin{equation}
a_{1}^{\mathrm{t-A}}=k_{1}t^{\ast }\eta .  \label{x150}
\end{equation}%
By applying $\delta $ on (\ref{x150}) we deduce
\begin{equation}
\delta a_{1}^{\mathrm{t-A}}=\frac{k_{1}}{4}\left( 4-D\right) \left(
3-D\right) \left( \partial ^{\mu }\partial ^{\rho }t_{\mu \rho }-\frac{1}{2}%
\Box t\right) \eta .  \label{x151}
\end{equation}%
In order to analyze the solution to the equation (\ref{x146}), we assume
that $a_{1}^{\mathrm{t-A}}$ of the form (\ref{x150}) generates a consistent $%
a_{0}^{\mathrm{t-A}}$. From (\ref{x151}) it follows that the corresponding $%
a_{0}^{\mathrm{t-A}}$ is linear in both $t_{\mu \nu \vert \alpha \beta }$ and $%
A_{\mu }$ and contains precisely one space-time derivative. Then, up to an
irrelevant divergence, $a_{0}^{\mathrm{t-A}}$ reads as%
\begin{equation}
a_{0}^{\mathrm{t-A}}=A_{\mu }m^{\mu }\left( \partial t\right) ,  \label{x152}
\end{equation}%
where $m^{\mu }\left( \partial t\right) $ is linear in the
first-order derivatives of $t_{\mu \nu \vert \alpha \beta }$. It is
simple to see that the most general form of $m^{\mu }\left( \partial
t\right) $ can be represented
like%
\begin{equation}
m^{\mu }\left( \partial t\right) =c_{3}\partial ^{\mu }t+c_{4}\partial
_{\rho }t^{\mu \rho },  \label{x153}
\end{equation}%
with $c_{3}$ and $c_{4}$ arbitrary, real constants, such that%
\begin{equation}
a_{0}^{\mathrm{t-A}}=A_{\mu }\left( c_{3}\partial ^{\mu }t+c_{4}\partial
_{\rho }t^{\mu \rho }\right) .  \label{x154}
\end{equation}%
By direct computation, from (\ref{x151}) and (\ref{x154}) we obtain
\begin{eqnarray}
\delta a_{1}^{\mathrm{t-A}}+\gamma a_{0}^{\mathrm{t-A}} &=&\partial _{\mu }%
\left[ \frac{k_{1}}{4}\left( 4-D\right) \left( 3-D\right) \left( \partial
_{\rho }t^{\mu \rho }-\frac{1}{2}\partial ^{\mu }t\right) \eta \right.
\notag \\
&&+\left( 4c_{3}+c_{4}\right) A^{\mu }\partial ^{\rho }\eta _{\rho \beta
\vert }^{\;\;\;\;\;\beta }  \notag \\
&&\left. +c_{4}A^{\alpha }\left( \partial ^{\mu }\eta _{\alpha \beta
\vert }^{\;\;\;\;\;\beta }-\partial ^{\beta }\eta _{\alpha \beta
\vert }^{\;\;\;\;\;\mu }\right) \right]  \notag \\
&&+\left( c_{3}+\frac{k_{1}}{8}\left( 4-D\right) \left( 3-D\right) \right)
\left( \partial ^{\mu }t\right) \partial _{\mu }\eta  \notag \\
&&+\left( c_{4}-\frac{k_{1}}{4}\left( 4-D\right) \left( 3-D\right) \right)
\left( \partial _{\rho }t^{\mu \rho }\right) \partial _{\mu }\eta  \notag \\
&&-\left( 4c_{3}+c_{4}\right) \left( \partial _{\mu }A^{\mu }\right)
\partial ^{\rho }\eta _{\rho \beta \vert }^{\;\;\;\;\;\beta }  \notag \\
&&-c_{4}\left( \partial _{\mu }A^{\alpha }\right) \left( \partial
^{\mu }\eta _{\alpha \beta \vert }^{\;\;\;\;\;\beta }-\partial
^{\beta }\eta _{\alpha \beta \vert }^{\;\;\;\;\;\mu }\right) .
\label{x155}
\end{eqnarray}%
The right-hand side of (\ref{x155}) reduces to a total derivative (as it is
required by the equation (\ref{x146})) if and only if the constants $k_{1}$,
$c_{3}$, and $c_{4}$ satisfy the equations%
\begin{eqnarray}
c_{3}+\frac{k_{1}}{8}\left( 4-D\right) \left( 3-D\right) &=&0,  \label{x156}
\\
c_{4}-\frac{k_{1}}{4}\left( 4-D\right) \left( 3-D\right) &=&0,  \label{x157}
\\
4c_{3}+c_{4} &=&0,  \label{x158} \\
c_{4} &=&0,  \label{x159}
\end{eqnarray}%
allowing only the vanishing solution
\begin{equation}
k_{1}=c_{3}=c_{4}=0.  \label{x160}
\end{equation}%
As a consequence, we find that
\begin{equation}
a_{1}^{\mathrm{t-A}}=0,  \label{x160a}
\end{equation}%
so $a^{\mathrm{t-A}}$ actually reduces to its component of antighost number
zero,
\begin{equation}
a^{\mathrm{t-A}}=a_{0}^{\mathrm{t-A}},  \label{xtv99}
\end{equation}%
which is subject to the `homogeneous' equation
\begin{equation}
\gamma a_{0}^{\mathrm{t-A}}=\partial _{\mu }\overset{\left( 0\right) }{m}%
^{\left( \mathrm{t-A}\right) \mu }.  \label{x161}
\end{equation}

There are two main types of solutions to this equation. The first type, to
be denoted by $a_{0}^{\prime \mathrm{t-A}}$, corresponds to $\overset{\left(
0\right) }{m}^{\left( \mathrm{t-A}\right) \mu }=0$ and is given by
gauge-invariant, nonintegrated densities constructed out of the original
fields and their space-time derivatives, which, according to (\ref{r79}),
are of the form
\begin{equation}
a_{0}^{\prime \mathrm{t-A}}=a_{0}^{\prime \mathrm{t-A}}\left( \left[
F_{\mu \nu }\right] ,\left[ F_{\mu \nu \lambda \vert \alpha \beta
\gamma }\right] \right) ,  \label{x161a}
\end{equation}%
up to the condition that they effectively describe cross-couplings between
the two types of fields and cannot be written in a divergence-like form.
Such a solution would produce vertices with more than two derivatives of the
fields and must be excluded since this disagrees with the hypothesis on the
maximum derivative order\footnote{%
If we however relax the derivative-order condition, then we can find
nonvanishing solutions of the type (\ref{x161a}).\ An example of a
possible solution is represented by the cubic vertex $a_{0}^{\prime
\mathrm{t-A}}=F_{\mu \nu \lambda \vert \alpha \beta \gamma }F^{\mu
\nu }F^{\alpha \beta }\sigma ^{\lambda \gamma }$.}
\begin{equation}
a_{0}^{\prime \mathrm{t-A}}=0.  \label{x161b}
\end{equation}

The second kind of solutions, to be denoted by $a_{0}^{\prime \prime \mathrm{%
t-A}}$, is associated with $\overset{\left( 0\right) }{m}^{\left( \mathrm{t-A%
}\right) \mu }\neq 0$ in (\ref{x161}), being understood that we discard the
divergence-like quantities and maintain the condition on the maximum
derivative order. In order to solve the equation
\begin{equation}
\gamma a_{0}^{\prime \prime \mathrm{t-A}}=\partial _{\mu }\overset{\left(
0\right) }{m}^{\left( \mathrm{t-A}\right) \mu },  \label{xww60}
\end{equation}%
we start from the requirement that $a_{0}^{\prime \prime \mathrm{t-A}}$ may
contain at most two derivatives. Then, $a_{0}^{\prime \prime \mathrm{t-A}}$
can be decomposed like
\begin{equation}
a_{0}^{\prime \prime \mathrm{t-A}}=\omega _{0}+\omega _{1}+\omega _{2},
\label{ww60}
\end{equation}%
where $\left( \omega _{i}\right) _{i=\overline{0,2}}$ contains $i$
derivatives. Due to the different number of derivatives in the components $%
\omega _{0}$, $\omega _{1}$, and $\omega _{2}$, the equation (\ref{xww60})
leads to three independent equations
\begin{equation}
\gamma \omega _{k}=\partial _{\mu }j_{k}^{\mu },\quad k=0,1,2.  \label{wwxy}
\end{equation}

For $k=0$ the equation (\ref{wwxy}) implies the necessary conditions $%
\partial _{\mu }\left( \partial \omega _{0}/\partial t_{\mu \nu \vert \alpha
\beta }\right) =0$ and $\partial _{\mu }\left( \partial \omega
_{0}/\partial A_{\mu }\right) =0$, whose solutions read as $\partial
\omega _{0}/\partial t_{\mu \nu \vert \alpha \beta }=k^{\mu \nu
\vert \alpha \beta }$ and $\partial \omega
_{0}/\partial A_{\mu }=k^{\mu }$, where $k^{\mu \nu \vert \alpha \beta }$ and $%
k^{\mu }$ are arbitrary, real constants. The last solutions provide
$\omega _{0}=k^{\mu \nu \vert \alpha \beta }t_{\mu \nu \vert \alpha
\beta }+k^{\mu }A_{\mu }$, so it does not describe
cross-interactions between $t_{\mu \nu \vert \alpha \beta }$ and
$A_{\mu }$ and can be made to vanish, $\omega _{0}=0$.\ For $k=1$
the equation (\ref{wwxy}) requires that
\begin{equation}
\partial _{\mu }\frac{\delta \omega _{1}}{\delta t_{\mu \nu \vert \alpha \beta }}%
=0,\quad \partial _{\mu }\frac{\delta \omega _{1}}{\delta A_{\mu }}=0,
\label{st1}
\end{equation}%
whose solutions are of the type\footnote{%
In general, the solution to the equation $\partial _{\mu }\left(
\delta \alpha /\delta t_{\mu \nu \vert \alpha \beta }\right) =0$ has
the form $\delta \alpha /\delta t_{\mu \nu \vert \alpha \beta
}=\partial _{\lambda }\partial _{\rho }M^{\mu \nu \lambda \vert
\alpha \beta \rho }+m^{\mu \nu \vert \alpha \beta }$, where the
functions $M^{\mu \nu \lambda \vert \alpha \beta \rho }$ display the
mixed symmetry of the curvature tensor and $m^{\mu \nu \vert \alpha
\beta }$ are some non-derivative constants with the mixed symmetry
$\left( 2,2\right) $. If we ask that $\alpha $\ comprises one
space-time derivative, then we must set $M^{\mu \nu \lambda \vert
\alpha \beta \rho }=0$ and $m^{\mu \nu \vert \alpha \beta }=0$,
which justifies the former solution from (\ref{st2}).}
\begin{equation}
\frac{\delta \omega _{1}}{\delta t_{\mu \nu \vert \alpha \beta }}=0,\quad \frac{%
\delta \omega _{1}}{\delta A_{\mu }}=\partial _{\nu }B^{\mu \nu },
\label{st2}
\end{equation}%
where $\delta \omega _{1}/\delta t_{\mu \nu \vert \alpha \beta }$
and $\delta \omega _{1}/\delta A_{\mu }$ denote the variational
derivatives of $\omega _{1}$. In the above the antisymmetric
functions $B^{\mu \nu }$ have no derivatives. Using (\ref{st2}), we
conclude that, up to an irrelevant divergence, $\omega _{1}$ is a
functions of $A_{\mu }$ with precisely one derivative. Such an
$\omega _{1}$ does not provide cross-interactions between $t_{\mu
\nu \vert \alpha \beta }$ and $A_{\mu }$, so we can take $\omega
_{1}=0$.

In the sequel we consider the equation (\ref{wwxy}) for $k=2$, which gives
the necessary conditions
\begin{equation}
\partial _{\mu }\frac{\delta \omega _{2}}{\delta t_{\mu \nu \vert \alpha \beta }}%
=0,\quad \partial _{\mu }\frac{\delta \omega _{2}}{\delta A_{\mu }}=0,
\label{st3}
\end{equation}%
with the solutions
\begin{equation}
\frac{\delta \omega _{2}}{\delta t_{\mu \nu \vert \alpha \beta
}}=\partial _{\lambda }\partial _{\rho }U^{\mu \nu \lambda \vert
\alpha \beta \rho },\quad \frac{\delta \omega _{2}}{\delta A_{\mu
}}=\partial _{\nu }\Phi ^{\mu \nu }. \label{st4}
\end{equation}%
Let $N$ be a derivation in the algebra of the fields and of their
derivatives that counts the powers of the fields and their derivatives,
defined by
\begin{eqnarray}
N &=&\sum\limits_{k\geq 0}\left( \left( \partial _{\mu _{1}}\cdots
\partial _{\mu _{k}}t_{\mu \nu \vert \alpha \beta }\right)
\frac{\partial }{\partial \left( \partial _{\mu _{1}}\cdots \partial
_{\mu _{k}}t_{\mu \nu \vert \alpha
\beta }\right) }\right.  \notag \\
&&\left. +\left( \partial _{\mu _{1}}\cdots \partial _{\mu _{k}}A_{\mu
}\right) \frac{\partial }{\partial \left( \partial _{\mu _{1}}\cdots
\partial _{\mu _{k}}A_{\mu }\right) }\right) .  \label{ww74}
\end{eqnarray}%
Then, it is easy to see that for every nonintegrated density $\chi $, we
have that
\begin{equation}
N\chi =t_{\mu \nu \vert \alpha \beta }\frac{\delta \chi }{\delta
t_{\mu \nu \vert \alpha \beta }}+A_{\mu }\frac{\delta \chi }{\delta
A_{\mu }}+\partial _{\mu }s^{\mu }.  \label{ww75}
\end{equation}%
If $\chi ^{\left( l\right) }$ is a homogeneous polynomial of order
$l>0$ in the fields $\left\{ t_{\mu \nu \vert \alpha \beta },A_{\mu
}\right\} $ and their derivatives, then $N\chi ^{\left( l\right)
}=l\chi ^{\left( l\right) }$. On account of (\ref{st4}) and
(\ref{ww75}), we find that
\begin{equation}
N\omega _{2}=\frac{1}{9}F_{\mu \nu \lambda \vert \alpha \beta \rho
}U^{\mu \nu \lambda \vert \alpha \beta \rho }+\frac{1}{2}F_{\mu \nu
}\Phi ^{\mu \nu }+\partial _{\mu }v^{\mu }.  \label{ww76a}
\end{equation}%
We expand $\omega _{2}$ like
\begin{equation}
\omega _{2}=\sum\limits_{l>0}\omega _{2}^{\left( l\right) },  \label{ww77}
\end{equation}%
where $N\omega _{2}^{\left( l\right) }=l\omega _{2}^{\left( l\right) }$,
such that
\begin{equation}
N\omega _{2}=\sum\limits_{l>0}l\omega _{2}^{\left( l\right) }.  \label{ww78}
\end{equation}%
Comparing (\ref{ww76a}) with (\ref{ww78}), we reach the conclusion that the
decomposition (\ref{ww77}) induces a similar decomposition with respect to $%
U^{\mu \nu \lambda \vert \alpha \beta \rho }$ and $\Phi ^{\mu \nu
}$, i.e.
\begin{equation}
U^{\mu \nu \lambda \vert \alpha \beta \rho
}=\sum\limits_{l>0}U_{\left( l-1\right) }^{\mu \nu \lambda \vert
\alpha \beta \rho },\quad \Phi ^{\mu \nu }=\sum\limits_{l>0}\Phi
_{\left( l-1\right) }^{\mu \nu }  \label{ww79}
\end{equation}%
Substituting (\ref{ww79}) into (\ref{ww76a}) and comparing the resulting
expression with (\ref{ww78}), we obtain that
\begin{equation}
\omega _{2}^{\left( l\right) }=\frac{1}{9l}F_{\mu \nu \lambda \vert
\alpha \beta
\rho }U_{\left( l-1\right) }^{\mu \nu \lambda \vert \alpha \beta \rho }+\frac{1}{%
2l}F_{\mu \nu }\Phi _{\left( l-1\right) }^{\mu \nu }+\partial _{\mu }\bar{v}%
_{(l)}^{\mu }.  \label{prform}
\end{equation}%
Introducing (\ref{prform}) in (\ref{ww77}), we arrive at
\begin{equation}
\omega _{2}=F_{\mu \nu \lambda \vert \alpha \beta \rho }\bar{U}^{\mu
\nu \lambda
\vert \alpha \beta \rho }+F_{\mu \nu }\bar{\Phi}^{\mu \nu }+\partial _{\mu }\bar{v%
}^{\mu },  \label{ww81}
\end{equation}%
where
\begin{equation}
\bar{U}^{\mu \nu \lambda \vert \alpha \beta \rho }=\sum\limits_{l>0}\frac{1}{9l}%
U_{\left( l-1\right) }^{\mu \nu \lambda \vert \alpha \beta \rho },\quad \bar{\Phi}%
^{\mu \nu }=\sum\limits_{l>0}\frac{1}{2l}\Phi _{\left( l-1\right) }^{\mu \nu
}.  \label{ww82}
\end{equation}%
By applying $\gamma $ on the relation (\ref{ww81}), after long and tedious
computation we infer that a necessary condition for the existence of
solutions to the equation $\gamma \omega _{2}=\partial _{\mu }j_{2}^{\mu }$
is that the functions $\bar{U}^{\mu \nu \lambda \vert \alpha \beta \rho }$ and $%
\bar{\Phi}^{\mu \nu }$ have the expressions
\begin{equation}
\bar{U}^{\mu \nu \lambda \vert \alpha \beta \rho }=C^{\mu \nu
\lambda \vert \alpha \beta \rho ;\sigma }A_{\sigma },\quad
\bar{\Phi}^{\mu \nu }=\bar{k}^{\mu \nu \rho ;\alpha \beta \vert
\sigma \lambda }\partial _{\rho }t_{\alpha \beta \vert \sigma
\lambda }, \label{st5}
\end{equation}%
where $C^{\mu \nu \lambda \vert \alpha \beta \rho ;\sigma }$ and
$\bar{k}^{\mu \nu \rho ;\alpha \beta \vert \sigma \lambda }$ are
non-derivative, real constants. The former constants exhibit the
mixed symmetry $\left( 3,3\right) $ in the indices $\mu \nu \lambda
\vert \alpha \beta \rho $ and are separately antisymmetric in
$\left\{ \alpha ,\beta ,\rho ,\sigma \right\} $. The quantities
$\bar{k}^{\mu \nu \rho ;\alpha \beta \vert \sigma \lambda }$ are
antisymmetric in the indices $\left\{ \mu ,\nu ,\rho \right\} $ and
display the mixed symmetry $\left( 2,2\right) $ with respect to
$\alpha \beta \vert \sigma \lambda $. Substituting (\ref{st5}) in
(\ref{ww81}) we get that
\begin{equation}
\omega _{2}=C^{\mu \nu \lambda \vert \alpha \beta \rho ;\sigma
}F_{\mu \nu
\lambda \vert \alpha \beta \rho }A_{\sigma }+\partial _{\rho }\left( F_{\mu \nu }%
\bar{k}^{\mu \nu \rho ;\alpha \beta \vert \sigma \lambda }t_{\alpha
\beta \vert \sigma \lambda }+\bar{v}^{\rho }\right) .  \label{st6}
\end{equation}%
As a consequence, the existence of a nontrivial $\omega _{2}$ is
conditioned by the existence of some pure constants $C^{\mu \nu
\lambda \vert \alpha \beta \rho ;\sigma }$ that must simultaneously
display the mixed symmetry $\left( 3,3\right) $ in their first six
indices and be
antisymmetric in the indices $\left\{ \alpha ,\beta ,\rho ,\sigma \right\} $%
. Because of the odd number of indices in $C^{\mu \nu \lambda \vert
\alpha \beta \rho ;\sigma }$, these constants can only be
constructed from the flat metric $\sigma ^{\mu \nu }$ and
Levi-Civita symbols $\varepsilon ^{\mu _{1}\cdots \mu _{j}}$. Due to
the identity $F_{\left[ \mu \right. \nu \lambda \vert \left. \alpha
\right] \beta \rho }\equiv 0$, the Levi-Civita symbols can be
contracted with $F_{\mu \nu \lambda \vert \alpha \beta \rho }$ on at
most three indices. On the other hand, the restriction $D\geq 5$ on
the space-time dimension requires Levi-Civita symbols with at least
five
indices, so $\varepsilon ^{\mu _{1}\cdots \mu _{j}}$ will contract with $%
F_{\mu \nu \lambda \vert \alpha \beta \rho }$ on at least four
indices, such that the corresponding $\omega _{2}$ will vanish
identically. In consequence, we can take
\begin{equation*}
C^{\mu \nu \lambda \vert \alpha \beta \rho ;\sigma }=0,
\end{equation*}%
which further leads to
\begin{equation}
a_{0}^{\prime \prime \mathrm{t-A}}=0.  \label{x164}
\end{equation}%
The relations (\ref{x161b}) and (\ref{x164}) show that
\begin{equation}
a_{0}^{\mathrm{t-A}}=0.  \label{x165}
\end{equation}%
By means of the results (\ref{3.25}), (\ref{x142}), (\ref{x143}), (\ref%
{x160a}), and (\ref{x165}) we arrive at
\begin{equation}
a^{\mathrm{t-A}}=0.  \label{x166}
\end{equation}

Finally, we focus on the solutions to the equation (\ref{uw2}). It is easy
to see that $a^{\mathrm{A}}$ can only reduce to its component of antighost
number zero
\begin{equation}
a^{\mathrm{A}}=a_{0}^{\mathrm{A}}\left( \left[ A_{\mu }\right] \right) ,
\label{vw1}
\end{equation}%
which is solution to the equation $sa^{\mathrm{A}}\equiv \gamma a_{0}^{%
\mathrm{A}}=\partial _{\mu }m_{0}^{\left( \mathrm{A}\right) \mu }$. It comes
from $a_{1}^{\mathrm{A}}=0$ and does not deform the gauge transformations,
but merely modifies the vector field action. The condition that $a_{0}^{%
\mathrm{A}}$ is of maximum derivative order equal to two is translated into
\begin{equation}
a_{0}^{\mathrm{A}}=c^{\prime \prime }\varepsilon ^{\mu \nu \lambda \beta
\rho }A_{\mu }F_{\nu \lambda }F_{\beta \rho },  \label{x167}
\end{equation}%
for $D=5$, with $c^{\prime \prime }$ an arbitrary, real constant. Putting
together the results deduced so far, we obtained that the first-order
deformation of the solution to the master equation for the theory (\ref{r1})
has the expression%
\begin{equation}
a=c^{\prime }t+c^{\prime \prime }\varepsilon ^{\mu \nu \lambda \beta \rho
}A_{\mu }F_{\nu \lambda }F_{\beta \rho }.  \label{x168}
\end{equation}

\subsection{Higher-order deformations}

Taking into account the equations (\ref{r61}), etc., we get that the
first-order deformation (\ref{x168}) is consistent to all orders in
the coupling
constant. Indeed, as $\left( S_{1},S_{1}\right) =0$, the equation (\ref{r61}%
), which describes the second-order deformation, is satisfied with
the choice
\begin{equation}
S_{2}=0,  \label{r99}
\end{equation}%
while the remaining higher-order equations are fulfilled for
\begin{equation}
S_{3}=S_{4}=\cdots =0.  \label{r100}
\end{equation}%
The fact that $a^{\mathrm{t-A}}=0$ shows there are no consistent
cross-couplings between the massless tensor field $t_{\mu \nu \vert
\alpha \beta } $ and the vector field $A_{\mu }$ complying with all
the hypotheses used in this paper.

\section{Conclusion}

To conclude with, in this paper we have investigated the couplings
between the massless tensor field with the mixed symmetry of the
Riemann tensor and the massless vector field by using the powerful
setting based on local BRST cohomology. Under the assumptions on
smoothness, locality, Lorentz covariance, and Poincar\'{e}
invariance of the deformations, combined with the requirement that
the interacting Lagrangian is at most second-order derivative, we
have proved that there are no consistent cross-interactions between
such fields. Our approach opens the perspective of investigating the
interactions between the tensor field $t_{\mu \nu \vert \alpha \beta }$ and one $%
p $-form ($p>1$) or, more general, between a tensor field with the mixed
symmetry $\left( k,k\right) $ and a $p$-form. These problems are under
consideration.

\section*{Acknowledgment}

The authors are partially supported by the European Commission FP6 program
MRTN-CT-2004-005104 and by the type A grant 304/2004 with the Romanian
National Council for Academic Scientific Research and the Romanian Ministry
of Education and Research.

\end{document}